\documentclass[prl,aps,superscriptaddress,twocolumn,bibnotes]{revtex4-2}
\usepackage{amsmath,amssymb}
\usepackage[colorlinks,bookmarks=false,citecolor=magenta,linkcolor=magenta,urlcolor=magenta]{hyperref}
\usepackage{tikz}
\usepackage{mathtools}
\usepackage{bm}
\usepackage{ragged2e}
\usepackage{graphicx}
\usepackage[symbol]{footmisc}
\usepackage[normalem]{ulem}

\begin{document}

	\title{Topological eigenvalues braiding and quantum state transfer near a third-order exceptional point}

	\author{He Zhang}
	\thanks{These authors contributed equally to this work.}
	\affiliation{Institute of Physics, Chinese Academy of Sciences, Beijing 100190, China}
	\affiliation{School of Physical Sciences, University of Chinese Academy of Sciences, Beijing 100190, China}

	\author{Tong Liu}
	\email{liutong181@mails.ucas.ac.cn}
	\thanks{These authors contributed equally to this work.}
	\affiliation{Institute of Physics, Chinese Academy of Sciences, Beijing 100190, China}
	\affiliation{Department of Microtechnology and Nanoscience, Chalmers University of Technology, 41296 Gothenburg, Sweden}

	\author{Zhongcheng Xiang}
	\affiliation{Institute of Physics, Chinese Academy of Sciences, Beijing 100190, China}
	\affiliation{School of Physical Sciences, University of Chinese Academy of Sciences, Beijing 100190, China}
	\affiliation{Beijing Academy of Quantum Information Sciences, Beijing 100193, China}
	\affiliation{Hefei National Laboratory, Hefei 230088, China}
	\affiliation{CAS Center of Excellence for Topological Quantum Computation, University of Chinese Academy of Sciences, Beijing 100190, China}
	\affiliation{Songshan Lake Materials Laboratory, Dongguan 523808, Guangdong, China}

	\author{Kai Xu}
	\affiliation{Institute of Physics, Chinese Academy of Sciences, Beijing 100190, China}
	\affiliation{School of Physical Sciences, University of Chinese Academy of Sciences, Beijing 100190, China}
	\affiliation{Beijing Academy of Quantum Information Sciences, Beijing 100193, China}
	\affiliation{Hefei National Laboratory, Hefei 230088, China}
	\affiliation{CAS Center of Excellence for Topological Quantum Computation, University of Chinese Academy of Sciences, Beijing 100190, China}
	\affiliation{Songshan Lake  Materials Laboratory, Dongguan 523808, Guangdong, China}

	\author{Heng Fan}
	\affiliation{Institute of Physics, Chinese Academy of Sciences, Beijing 100190, China}
	\affiliation{School of Physical Sciences, University of Chinese Academy of Sciences, Beijing 100190, China}
	\affiliation{Beijing Academy of Quantum Information Sciences, Beijing 100193, China}
	\affiliation{Hefei National Laboratory, Hefei 230088, China}
	\affiliation{CAS Center of Excellence for Topological Quantum Computation, University of Chinese Academy of Sciences, Beijing 100190, China}
	\affiliation{Songshan Lake Materials Laboratory, Dongguan 523808, Guangdong, China}

	\author{Dongning Zheng}
	\affiliation{Institute of Physics, Chinese Academy of Sciences, Beijing 100190, China}
	\affiliation{School of Physical Sciences, University of Chinese Academy of Sciences, Beijing 100190, China}
	\affiliation{Hefei National Laboratory, Hefei 230088, China}
	\affiliation{CAS Center of Excellence for Topological Quantum Computation, University of Chinese Academy of Sciences, Beijing 100190, China}
	\affiliation{Songshan Lake  Materials Laboratory, Dongguan 523808, Guangdong, China}

	\begin{abstract}
	
	Non-Hermitian systems exhibit a variety of unique features rooted in the presence of exceptional points (EP). 
	The distinct topological structure in the proximity of an EP gives rise to counterintuitive behaviors absent in Hermitian systems, which emerge after encircling the EP either quasistatically or dynamically. 
	However, experimental exploration of EP encirclement in quantum systems, particularly those involving high-order EPs, remains challenging due to the difficulty of coherently controlling more degrees of freedom.
	In this work, we experimentally investigate the eigenvalues braiding and state transfer arising from the encirclement of EP in a three-dimensional non-Hermitian quantum system using superconducting circuits.
	We characterize the second- and third-order EPs through the coalescence of eigenvalues.
	Then we reveal the topological structure near the EP3 by quasistatically encircling it along various paths with three independent parameters, which yields the eigenvalues braiding described by the braid group $B_3$.
	Additionally, we observe chiral state transfer between three eigenstates under a fast driving scheme when no EPs are enclosed, while time-symmetric behavior occurs when at least one EP is encircled.
	Our findings offer insights into understanding non-Hermitian topological structures and the manipulation of quantum states through dynamic operations. 

	\end{abstract}

	\maketitle

	\section{Introduction}

	The exploration of non-Hermitian systems has uncovered a wide range of intriguing phenomena, such as unidirectional invisibility~\cite{PhysRevLett.106.213901,Feng2012}, perfect absorption~\cite{PhysRevLett.112.143903,PhysRevLett.122.093901}, lasing effects~\cite{Peng2014,Feng2014,Hodaei2014}, frequency combs~\cite{Wang2024}, and unconventional beam dynamics~\cite{PhysRevLett.100.103904}.
	These findings are fostered by the presence of non-Hermitian degeneracies known as $k$th-order exceptional points (EP$k$s), where $k$ eigenvalues and their corresponding eigenstates both simultaneously coalesce.
	The nontrivial topological structure near an EP within the Riemann manifold leads to unique behaviors that have no counterpart in Hermitian systems.

	In an $N$-dimensional non-Hermitian system, traversing a directional control loop parametrized by $s$ around an EP in a quasistatic manner results in a permutation of $N$ complex eigenvalues $\{\lambda_i(s)\}$~\cite{Kato1995,Ashida2020}.
	The trajectories of eigenvalues in the three-dimensional space $(\mathrm{Re}[\lambda], \mathrm{Im}[\lambda], s)$ resemble $N$ intertwined strands of a braid.
	For the EPs in a two-dimensional non-Hermitian system, the topology of control loops in a two-dimensional parameter space can be classified by the group $\mathbb{Z}$, which also corresponds to the braid degree of two eigenvalues along the loop~\cite{Wang2021}.

	To predict the evolution of $N$ eigenvalues, however, it is necessary to account for the topological structure of the space $\mathcal G_N$, excluding degenerate points in a higher-dimensional parameter space of $2(N-1)$-dimensions~\cite{Guria2024}.
	A bijection can be established between the fundamental group of the space $\mathcal G_N$ and the Artin braid group $B_N$~\cite{PhysRevB.101.205417,PhysRevB.103.155129,PhysRevLett.126.010401}, which is a non-Abelian group for $N>2$.
	In the case of $N=3$, the full parameter space is four-dimensional, with all EPs forming a two-dimensional space isomorphic to a cone of the trefoil knot $\mathcal K\times\mathbb{R}_{>0}$ where $\mathcal K$ denotes the trefoil knot~\cite{Patil2022}.
	To locate the EP3 and ascertain the braid relation, four independent parameters must be adjusted, which demands substantial experimental effort~\cite{Patil2022}. 
	Recent studies suggest that additional symmetries can lessen the degrees of freedoms required to observe high-order EPs~\cite{Tang2023,Wu2024}.
	Therefore, it remains to be explored whether digesting the topology of encircling EPs and reproducing the associated braid group $B_N$ still necessitates $2(N-1)$ parameters when such symmetries are present.

	Another noteworthy phenomenon is the mode switch among eigenstates induced by the dynamical encirclement of EPs. 
	Instead of the eigenstate exchange observed in quasistatic encirclement, non-adiabatic transitions during the dynamical cycle lead to the breakdown of the adiabatic theorem~\cite{Uzdin2011,PhysRevA.88.010102,PhysRevA.92.052124,PhysRevLett.118.093002,PhysRevA.102.040201}. 
	The final state is determined by the encirclement direction and the location of the starting point, irrespective of the initial state.
	While such novel behaviors have been demonstrated in classical systems~\cite{Xu2016,Doppler2016,Yoon2018,PhysRevX.8.021066,Zhang2019,Schumer2022,Nasari2022}, extending this topological control to quantum systems demands both high controllability and long coherence time.
	Although recent studies have realized the dynamical encirclement of a second-order EP in real quantum systems~\cite{PhysRevLett.126.170506,PhysRevLett.128.160401}, the exploration of higher-order EP remains elusive due to the challenge of precisely controlling more time-dependent parameters.
	Moreover, the chiral state transfer can occur even in the absence of encircled EPs~\cite{PhysRevA.96.052129, Nasari2022,PhysRevLett.128.160401}, which has yet to be observed in high-dimensional non-Hermitian systems.

	In this work, we investigate eigenvalues braiding and state transfer in a three-dimensional non-Hermitian quantum system comprising a transmon coupled with a microwave resonator.
	Leveraging the multilevel structure of transmon, we approximate it as a three-level quantum system and achieve tunable coupling between the transmon and the resonator.
	Dissipation is introduced by the resonator, which undergoes much faster photon decay compared to the transmon~\cite{PhysRevX.4.041010,PhysRevA.91.043846,PhysRevLett.121.060502}.
	We identify the EP2s and EP3 in this system by the coalescence of eigenvalues.
	We develop a mapping between the control loop and the braid of eigenvalues, and illustrate how to employ this mapping to generate the complete braid group $B_3$ with three parameters.
	We reveal that both time-symmetric and chiral state transfers can appear by choosing different paths for the dynamical encirclement of EPs, which can be implemented through a rapid driving scheme.
	The chiral state transfer manifests only when no EPs are surrounded.

	\section{Results}

	\begin{figure}[ht]
		\centering
		\includegraphics[width=\linewidth]{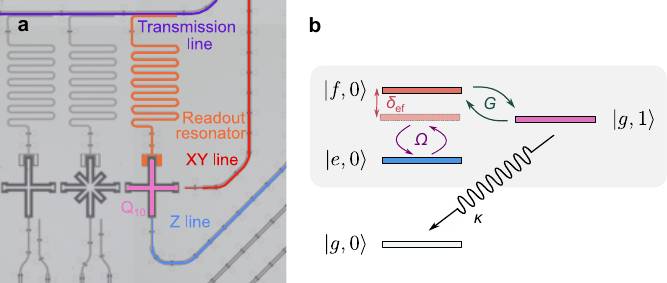}
		\caption{\quad Three-level non-Hermitian systems. (a) The optical picture of the ten-qubit superconducting quantum processor with highlighting circuit elements of a transmon and its attached resonator. 
		Scale bar at the bottom, 0.2 mm.
		(b) The non-Hermitian system is composed of the states $|e,0\rangle$, $|f,0\rangle$, and $|g,1\rangle$, with dissipation irreversibly driving the transition from $|g,1\rangle$ to $|g,0\rangle$. 
		}
		\label{fig:chip}
	\end{figure}

	\begin{figure*}[ht]
		\centering
		\includegraphics[width=\textwidth]{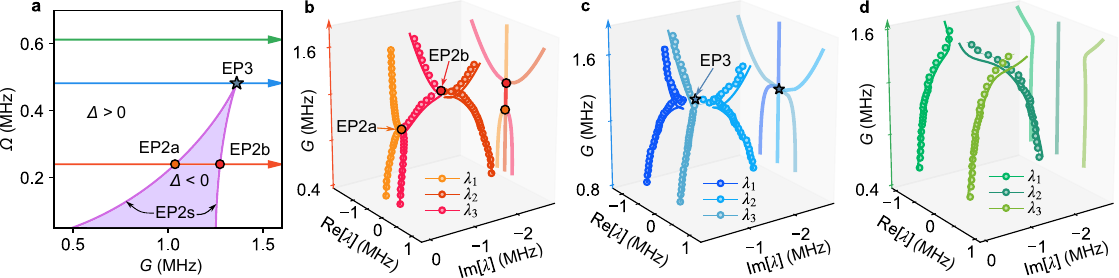}
		\caption{\quad Determine EP2s and EP3.
		(a) The phase diagram in the first quadrant of the plane $\delta_{ef}=0$. 
		Purple arcs represent EP2s and the blue star denotes the EP3.
		The colored region surrounded by EP2s represents the anti-PT-symmetry-broken phase with discriminant $\Delta<0$.
		Red, blue, and red arrows represent specific routines.
		(b)-(d) Retrieved eigenvalues corresponding to the routines in (a). Dots are experimental data and solid lines are theoretic results.
		The standard error of the mean in the measured data is no larger than the size of the plotted points.
		The solid lines on the projected planes show the real parts of the eigenvalues.}
		\label{fig:EP}
	\end{figure*}

	\subsection{Symmetry-enabled realization of high-order EPs}

	We construct a non-Hermitian system on a superconducting quantum processor with a transmon and a capacitively coupled microwave resonator.
	Figure~\ref{fig:chip}(a) illustrates the processor utilized in our experiment, which contains a one-dimensional array of 10 frequency-tunable transmon elements.
	We only exploit $Q_{10}$ in the experiment, while the others are detuned from the experimental frequency using magnetic flux bias to reduce crosstalk. 
	The transmon $Q_{10}$ can be regarded as a qutrit comprising its ground state $|g\rangle$, the first excited state $|e\rangle$ and the second excited state $|f\rangle$, with an energy relaxation time of $T_e=48$ $\mu$s and $T_f=35$ $\mu$s. 
	The initial state preparation and readout are operated at the sweet spot frequencies $\omega_{ge}=2\pi\times 5.684$ GHz and $\omega_{ef}=\omega_{ge} + \alpha = 2\pi\times 5.431$ GHz with an anharmonicity $\alpha/2\pi = - 253$ MHz. 
	The transitions between $g$-$e$ and $e$-$f$ are individually addressed by different modulated microwave pulses generated by arbitrary waveform generators (AWGs).
	The transmon is coupled with the strength $J/2\pi=48$ MHz to a readout resonator of frequency $\omega_r/2\pi=6.697$ GHz.
	The photon decay rate $\kappa$ of resonator is measured to be 5 MHz by the ac Stark effect~\cite{PhysRevA.76.042319}.

	Although the coupling strength $J$ is fixed by the geometry of the device, cavity-assisted Raman process enables a controllable effective coupling $G$ between $|g,1\rangle$ and $|f,0\rangle$ by employing a coherent microwave drive where $|s,n\rangle$ denotes the product state of the transmon in state $|s\rangle$ and the resonator in the $n$ photon Fock state $|n\rangle$~\cite{PhysRevX.4.041010,PhysRevA.91.043846,PhysRevLett.121.060502}.
	By applying an additional Rabi drive between the $|e,0\rangle$ and $|f,0\rangle$ states, the dynamics of the coupled system in the subspace spanned by the $|e,0\rangle$, $|f,0\rangle$, and $|g,1\rangle$ states is governed by the non-Hermitian Hamiltonian
	\begin{equation}
		H = \begin{pmatrix}
			-\delta_{ef} & \Omega & 0 \\
			\Omega & 0 & G \\
			0 & G & -i\kappa/2
		\end{pmatrix},\label{eq:NH-Hamiltonian}
	\end{equation}
	where $\delta_{ef}$ is the drive detuning, accounting for the ac Stark shift induced by the $f0$-$g1$ drive, and $\Omega$ is the drive amplitude between the $|e,0\rangle$ and $|f,0\rangle$ states.
	For simplicity, $\Omega$ and $G$ are both tuned to be real. 
	The ac Start shift for the $|g,1\rangle$ state induced by the $f0$-$g1$ drive is consistently compensated for in the experiment.

	The non-Hermitian Hamiltonian in Eq.~(\ref{eq:NH-Hamiltonian}) can be parameterized with three real independent variables $(\delta_{ef}, \Omega, G)$, which can be accurately mapped onto the raw instrument parameters following a series of calibrations prior to the experiment~\cite{sm}.
	However, the existence of EP$k$s generally necessitate $2(k-1)$ constraints implying that at least four free parameters are required to investigate EP3s~\cite{Tang2020,Patil2022,Tang2023}.
	To resolve this discrepancy, we mention that the Hamiltonian possesses a pseudo-chirality at $\delta_{ef}=0$ which lowers the number of constraints to two~\cite{PhysRevLett.127.186601,PhysRevLett.127.186602}.
	Therefore, we can observe an isolated EP3 at the two-dimensional surface $\delta_{ef}=0$ in the parameter space.
	To explicitly demonstrate it, we consider the characteristic polynomial of the Hamiltonian $\varphi_\lambda(H)=\lambda^3+(\delta_{ef} + i\kappa/2)\lambda^2 - (G^2 + \Omega^2 - i\kappa\delta_{ef}/2)\lambda - (G^2\delta_{ef} + i\kappa\Omega^2/2) $.
	The roots of this cubic polynomial yield three eigenvalues $\lambda_1$, $\lambda_2$, and $\lambda_3$.
	All of them can be expressed as $(-\frac{p}{2}+\sqrt \Delta)^{1/3}\alpha + (-\frac{p}{2} - \sqrt\Delta)^{1/3}\alpha^* - (\delta_{ef}+i\frac{\kappa}{2})/3$ where $\alpha$ takes the value from $\{1,e^{i2\pi/3},e^{i4\pi/3}\}$ using Cardano's formula.
	Here, $\Delta$ denotes the discriminant of $\varphi_\lambda(H)$, and $p$ is a function about $\delta_{ef}$, $\Omega$ and $G$.
	The square root of $\Delta$ indicates that EP2s emerge at $\Delta=0$ and coalesce into an EP3 when $p=0$ as well.
	We plot the EP2s and EP3 in the first quadrant of the plane $\delta_{ef}=0$ in Fig.~\ref{fig:EP}a.
	The boundary lines of the shaded region $\Delta < 0$ consist of two branches of EP2s, and EP3 is located at the cusp of order-2 exceptional arcs.
	In addition, the system preserves an anti-PT symmetry at $\delta_{ef}=0$~\cite{Peng2016,Choi2018,Li2019,PhysRevLett.125.147202,Bergman2021}.
	The sign change of $\Delta$ characterizes an anti-PT-symmetry phase transition.
	In the symmetry-unbroken regime ($\Delta < 0$), all eigenvalues are purely imaginary, leading to exponential decay dynamics.
	Spontaneous symmetry breaking occurs at EP2s $(\Delta=0)$, where two eigenvalues degenerate.
	When $\Delta>0$, the eigenvalues become complex and result in damped oscillations. 

	We experimentally examine EPs along three distinct routines that traverse exceptional arcs with two, one and zero crossings on the plane $\delta_{ef}=0$, as shown in Fig.~\ref{fig:EP}a.
	We fix the Rabi driving amplitude $\Omega$ between the $|e,0\rangle$ and $|f,0\rangle$ states while gradually increasing the amplitude of $f0$-$g1$ driving pulse.
	Simultaneously, we adjust the driving frequency to maintain zero detuning.
	For each point on the trajectories, we prepare the initial state as $(|e,0\rangle + |f,0\rangle)/\sqrt 2$ and measure the evolution of the probabilities for the three qutrit states $|g\rangle$, $|e\rangle$, and $|f\rangle$.
	The retrieved eigenvalues are obtained from the non-Hermitian Hamiltonian using the extracted experiment parameters $(\delta_{ef}^{\mathrm{exp}}, \Omega^{\mathrm{exp}}, G^{\mathrm{exp}})$.
	These parameters are determined by fitting the population dynamics of three states simultaneously. 

	Figure~\ref{fig:EP}b-d present the retrieved eigenvalues corresponding to the three specific routines shown in Fig.~\ref{fig:EP}a, respectively.
	All experiment data align well with the theoretic predictions.
	In Fig.~\ref{fig:EP}b, one eigenvalue remains purely imaginary while the other two are symmetrically mirrored about the plane $\mathrm{Re}[\lambda]=0$ as $G$ increases, before converging at the first EP2 (EP2a) in the left branch.
	Up crossing EP2a, all eigenvalues turn imaginary until they reach another EP2 (EP2b) in the right branch.
	Passing through the symmetry-unbroken region, two eigenvalues become complex again.
	The real part of the eigenvalues undergoes pitchfork bifurcations at two EP2s, as depicted on the projected plane in Fig.~\ref{fig:EP}b. 
	Notably, the symmetric eigenvalues after crossing EP2b are distinct from the pair before crossing EP2a as illustrated in Fig.~\ref{fig:EP}b, which gives rise to two types of eigenvalues braiding when encircling EP2s in different branches.
	Next, we follow the blue routine where $\Omega/\kappa=3^{-3/2}/2$ and encounter the EP3 at $G/\kappa=(2/3)^{3/2}/2$~\cite{sm}.
	Figure~\ref{fig:EP}c plots the eigenvalues in the vicinity of EP3 where all three eigenvalues coalesce, confirming the existence of EP3.
	The inner projected plane also shows the merging of two bifurcations points into one.
	In contrast, for the green routine without intersecting the region where $\Delta < 0$, no eigenvalues coalescence is observed in Fig.~\ref{fig:EP}d as anticipated.

	\begin{figure*}
		\centering
		\includegraphics[width=0.95\linewidth]{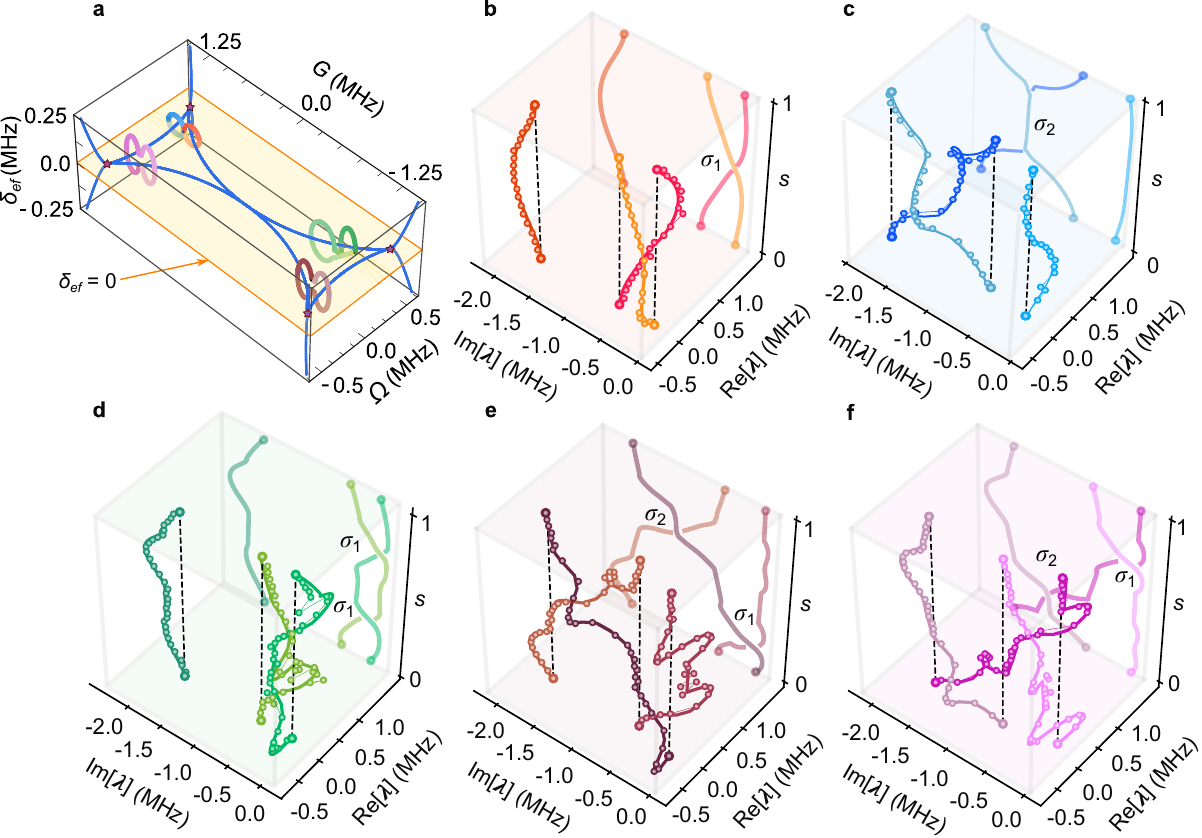}
		\caption{\quad Braid of eigenvalues. 
		(a) EPs and control loops. 
		Blue solid lines represent EPs.
		Red, blue, green, brown, and purple loops represent the control loops around EPs. 
		Each loop is parameterized as $l(s)$ and completes a full cycle as $s$ increases from 0 to 1.
		The yellow plane denotes the surface $\delta_{ef}=0$, and the stars mark the four EP3s.
		(b)-(f) Braid of eigenvalues corresponding to the five control loops in (a).
		Dots represent the experiment data and solid lines indicate theoretic results.
		The standard error of the mean in the measured data is no larger than the size of the plotted points.
		The projected planes offer a front-view perspective of braids.
		}
		\label{fig:braid}
	\end{figure*}

	\subsection{Complex eigenvalues braiding}
	
	After identifying EPs on the surface $\delta_{ef}=0$, we investigate the behavior of eigenvalues by varying the parameters of $H$ around a loop in the parameter space.
	Contrary to the requirement of a four-dimensional parameter space~\cite{Patil2022}, we demonstrate that three independent parameters are sufficient to realize the entire braid group $B_3$.
	Figure~\ref{fig:braid}a illustrates EPs within the three-dimensional parameter space.
	We focus on the EPs depicted by the blue solid lines, which continue infinitely in the parameter space as $|\delta_{ef}|$ increases, rather than constituting a closed knot~\cite{Patil2022}. 
	Two additional exceptional lines, defined by $\{(\delta_{ef},\Omega, G)|\Omega=0,G=\pm \kappa/4\}$, are not joined with the main framework of EPs and are not shown in the figure, as they do not impact our results.
	Owing to the absence of the coupling between $|e,0\rangle$ and $|g,1\rangle$, the eigenvalues are invariant regardless of the phase variations or the signs of $\Omega$ and $G$ (See Methods).
	Consequently, the EPs shown in the first quadrant of the surface $\delta_{ef}=0$ (Fig.~\ref{fig:EP}a) extend to four quadrants in Fig.~\ref{fig:braid}a.

	\begin{figure*}
		\centering
		\includegraphics[width=\linewidth]{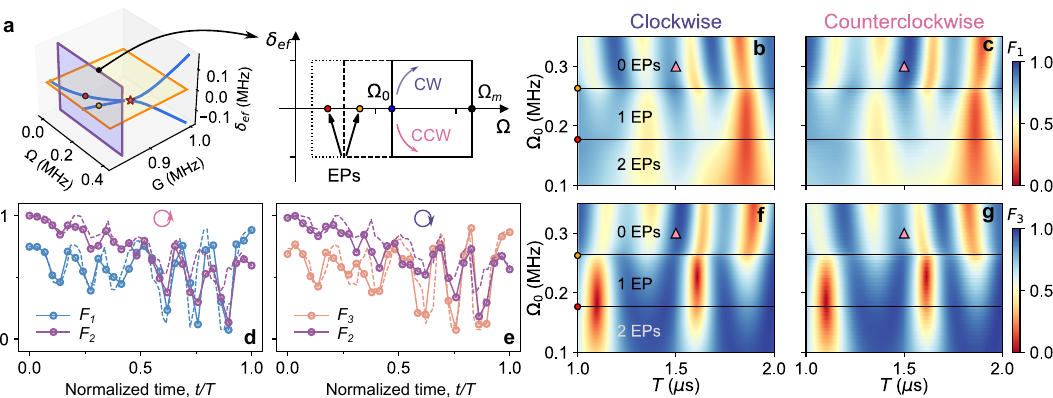}
		\caption{\quad Dynamically encircle EPs. (a) The sketch of encirclement loops on the plane $G=0.845$ MHz. 
		By adjusting the initial point $(\Omega_0, 0)$, the number of EPs encircled by the loop can be reduced from two to zero. 
		The encirclement direction is either clockwise (CW) or counterclockwise (CCW).
		(b)(c) and (f)(g) The numerical results of state overlap $F_j= |\langle\psi_j|\psi(T)\rangle|^2$ over the encirclement period $T$ and $\Omega_0$ where $|\psi(t)\rangle$ is the normalized state at time $t$ and $|\psi_j\rangle$ is the $j$-th normalized right eigenstate of initial Hamiltonian with $j=1,2$ and 3.
		The initial state is prepared at the right eigenstate $|\psi_2\rangle$.
		(d)(e) The state overlaps during the evolution using the parameter marked by the pink triangle in (b)(c) and (f)(g).
		The circles represent the experiment data and the dashed lines represent the numerical simulation results.
		The standard error of the mean in the measured data is no larger than the size of the plotted points.}
		\label{fig:chiral}
	\end{figure*}

	To observe the spectral flow, we select five control loops which share a common base point severing as both start and end points, as shown in Fig.~\ref{fig:braid}a.
	For clarity, the control loops in green, purple, and brown are symmetrically mirrored in the other quadrants.
	The five loops belonging to different homotopy classes give rise to different braids of eigenvalues in Fig.~\ref{fig:braid}b-f, where the evolution of each eigenvalue resembles a strand.
	The red and blue loops encircling distinct branches of EP2s on the surface $\delta_{ef}=0$ generate elements $\sigma_1$ and $\sigma_2$ of the braid group $B_3$, respectively.
	By concatenating these loops, we can derive any element in $B_3$, such as $\sigma_1^2$, $\sigma_2\sigma_1$, and $\sigma_1\sigma_2$ (Fig.~\ref{fig:braid}d-f).
	As illustrated in Fig.~\ref{fig:braid}e and f, the braids formed by concatenating $\sigma_1$ and $\sigma_2$ in opposite orders are not equivalent, verifying that $B_3$ is a non-Abelian group.
	Apart from the decomposition into red and blue loops, the purple or brown loop can also be interpreted as a concatenation of loops encircling the other two exceptional arcs with $\delta_{ef}\ne 0$~\cite{sm}.
	The equivalence of two concatenations leads to the braid identity $\sigma_1\sigma_2\sigma_1=\sigma_2\sigma_1\sigma_2$, which is also known as the Yang-Baxter equation~\cite{Yang1991}.
	From another prospective, the purple and brown loops can be continuously deformed into the loops around EP3 without concatenation, producing a full braid of three eigenvalues in one step.
	Similarly, the red and blue loops can be viewed as encircling EP3 within a plane which separates one exceptional arc from the other three, producing a braid between two eigenvalues.
	These results indicate the anisotropy of EP3 stemming from the novel geometry of the space $\mathcal G_N$~\cite{Tang2020}.

	We mention that such a topological structure close to the EP3 cannot be fully characterized by the discriminant number $\nu=\sum_{i\ne j}\nu_{ij}$ where $\nu_{ij}$ is the vorticity invariant between eigenvalues $\lambda_i$ and $\lambda_j$~\cite{PhysRevLett.126.086401,PhysRevLett.127.034301}.
	It can be verified that $\nu=-1$ for the loops shown in Fig.~\ref{fig:braid}b-c and $\nu=-2$ for the loops shown in Fig.~\ref{fig:braid}d-f. 
	Another promising topological invariant is the knot invariant, obtained by connecting the corresponding ends of the braid to form a knot or link~\cite{PhysRevLett.126.010401,Wang2021}.
	However, different braids may yield the same knot or link, as seen in Fig.~\ref{fig:braid}b-c, where both braids give rise to the same unlink, making them indistinguishable by knot invariants.
	Therefore, the braid group is a more fundamental tool for a comprehensive understanding of the topological structure near a high-order EP.

	\subsection{Chiral quantum state transfer}

	We now turn to the case of dynamically varying parameters along a cycle, where only one eigenstate dominates at the end of the evolution.
	In a two-dimensional non-Hermitian system, if a loop produces one eigenstate and its reverse yields the other one, it is called chiral state transfer; otherwise, it is termed time-symmetric state transfer. 
	In higher-dimensional systems, this process is more intricate due to the presence of more eigenstates and EPs.
	To investigate the impact of EP on state transfer, we confine the dynamical loops within the plane $G=0.845$ MHz which intersects with the exceptional arcs twice creating two EPs, as shown in Fig.~\ref{fig:chiral}a.
	Each loop is a rectangle with a height of $2a$ and a width of $\Omega_m-\Omega_0$ where $a/2\pi=5$ MHz is the maximum value of detuning and $\Omega_m=5$ MHz is the fixed maximum value of $\Omega$.
	The four edges of the rectangle are swept consecutively in either a clockwise (CW) or counterclockwise (CCW) direction at different constant velocities, ensuring that each edge is traversed in $T/4$ time, where $T$ is the total period.
	The starting point of the loop is situated at $(\Omega_0, 0)$.
	By increasing the value of $\Omega_0$, the loop is squeezed and the number of surrounding EPs decreases from two to zero.

	Since the eigenstate is expressed in the basis involving the resonator's state, we develop an approach to prepare a specific initial state by successively applying a $g$-$e$ pulse, $e$-$f$ pulse and a $f0$-$g1$ pulse~\cite{sm}. 
	To characterize the final state, we first perform standard quantum state tomography between the states $|e,0\rangle$ and $|f,0\rangle$.
	Then we continue the evolution without the $e$-$f$ Rabi drive to extract the information about the $|g,1\rangle$~\cite{sm}.
	During the evolution, we dynamically adjust the phase of the $e$-$f$ pulse to correct the unwanted phase shifts arising from changes in detuning~\cite{sm}.

	To examine the exchange among the eigenstates along different loops, we prepare the initial state as the right eigenstate $|\psi_2\rangle$ and allow the system to evolve according to the loop, where $|\psi_j\rangle$ represents the $j$-th normalized right eigenstate associated with the eigenvalue $\lambda_j$.
	Figure~\ref{fig:chiral}b, c, f and g show the simulation results of overlap $F_j = |\langle\psi_j|\psi(T)\rangle|^2$ between the normalized final state $|\psi(T)\rangle$ and the remaining right eigenstates $|\psi_1\rangle$ and $|\psi_3\rangle$.
	Different from the adiabatic limit ($T\rightarrow\infty$), the state transfer is strongly influenced by both the period $T$ and the position of the starting point~\cite{PhysRevLett.128.160401}.
	For the loops encompassing two EPs, the final state always transfers to the eigenstate $|\psi_3\rangle$ at some specific times regardless of the direction.
	Similar time-symmetric behaviors are also observed for the loops enclosing one EP.
	However, chiral state transfer occurs when no EPs are surrounded by the loops.
	As indicated by the pink triangles in Fig.~\ref{fig:chiral}b, c, f, and g, the CW loop yields the transfer from $|\psi_2\rangle$ to $|\psi_3\rangle$, while the CCW loop causes the transfer from $|\psi_2\rangle$ to $|\psi_1\rangle$.
	Figure~\ref{fig:chiral}d-e plot the evolution of state overlap $F_j$ over time in experiments using the parameters marked by the pink triangle.
	The experimental results are consistent with the numerical results.
	The oscillations in the overlaps during the evolution underline the importance of selecting an appropriate evolution time.

	\section{Conclusion and discussion}
	We realize a non-Hermitian system containing an EP3 using superconducting circuits.
	We demonstrate that the EP3 is the intersection of multiple exceptional arcs.
	By identifying the exceptional arcs in the three-dimensional parameter space, we illustrate the generation of braid group $B_3$ by concatenating different classes of control loops.
	We also explore the chiral state transfer between the eigenstates over a relatively short time by dynamically encircling the vicinity of the EP without enclosing it.
	Compared to the adiabatic approach, shortening the evolution time suppresses the decoherence effects of the device.
	We mention that the state transfer vanishes when the starting point is set at $(\Omega_m,0)$, which highlights the role of the starting position~\cite{PhysRevX.8.021066}.

	The study of non-Hermitian systems can be greatly advanced through the flexibility and scalability of superconducting circuits.
	Beyond controlling coherent parameters, the dissipation rate can be enhanced by incorporating a Purcell filter without degrading the coherence of transmon~\cite{PhysRevLett.112.190504,PhysRevA.92.012325}, or dynamically tuned via a normal metal-insulator-superconductor tunnel junction~\cite{Tan2017,PhysRevB.100.134505}.
	By coupling multiple transmons, the composite system facilitates the exploration of entanglement generation near higher-order EPs~\cite{PhysRevLett.131.100202} or enable Bell-state transfer between qubits~\cite{PhysRevLett.133.070403}.

	\section{Methods}

	\subsection{Symmetries of non-Hermitian Hamiltonian}
	Pseudo-chirality and anti-PT-symmetry are defined as $U_\mathrm{psCh}HU_\mathrm{psCh}^{-1} = -H^\dagger$ and $U_\mathrm{APT}HU_\mathrm{APT}^{-1} = -H^*$, respectively, where $U_\mathrm{psCh}$ and $U_\mathrm{APT}$ are two unitary operators and $H$ represents the non-Hermitian Hamiltonian.
	The anti-PT-symmetry can be derived from the PT-symmetry by substituting $H\rightarrow iH$.
	By choosing
	\begin{equation}
		U_\mathrm{psCh} = U_\mathrm{APT} = \begin{pmatrix}
			1 & 0 & 0 \\
			0 & -1 & 0 \\
			0 & 0 & 1
		\end{pmatrix},
	\end{equation}
	we can verify that the Hamiltonian in Eq.~(\ref{eq:NH-Hamiltonian}) preserves both symmetries when $\delta_{ef}=0$.
	For a nonzero $\delta_{ef}$, the eigenvalues of the Hamiltonian remain unchanged under the transformations $\Omega\rightarrow\Omega e^{i\phi}$ and $G\rightarrow Ge^{i\theta}$ by noticing that $U_1(\phi)H(\delta_{ef}, \Omega, G)U_1^{-1}(\phi) = H(\delta_{ef}, \Omega e^{i\phi}, G)$ and $U_2(\theta)H(\delta_{ef},\Omega, G)U_2^{-1}(\theta)=H(\delta_{ef},\Omega, Ge^{i\theta})$ 
	where 
	\begin{equation}
		U_1(\phi) = \begin{pmatrix}
			e^{i\phi} & 0 & 0 \\
			0 & 1 & 0 \\
			0 & 0 & 1
		\end{pmatrix},
	\end{equation}
	and
	\begin{equation}
		U_2(\theta) = \begin{pmatrix}
			1 & 0 & 0 \\
			0 & 1 & 0 \\
			0 & 0 & e^{i\theta}
		\end{pmatrix}.
	\end{equation}


%

	\section{Acknowledgements}
	This work was supported by the National Natural Science Foundation of China (Grant Nos.~12204528, 92265207, T2121001, 92065112), the Innovation Program for Quantum Science and Technology (Grant No.~2021ZD0301800), and and Beijing Natural Science Foundation (Grant No.~Z200009).
	This work also was supported by the Micro/nano Fabrication Laboratory of Synergetic Extreme Condition User Facility (SECUF). 
	Devices were made at the Nanofabrication Facilities at the Institute of Physics, CAS in Beijing.

\end{document}